\documentclass[12pt]{article}
\pdfoutput=1

\usepackage[normalem]{ulem}
\usepackage[utf8]{inputenc}
\usepackage[left=2.55cm, right=2.55cm, top=2.55cm, bottom=2.55cm]{geometry}
\usepackage{amsmath,amssymb}
\usepackage{slashed}
\usepackage{xcolor}
\usepackage{graphicx}
\usepackage{url}
\usepackage{cancel}
\usepackage{cite}
\usepackage[colorlinks=true,allcolors=darkpurple,pdfborder={0 0 0},linktocpage=false]{hyperref}
\usepackage{tabularx,booktabs}
\usepackage{multicol}
\usepackage{units}
\usepackage{xspace}
\usepackage[labelfont=bf]{caption}
\usepackage{tikz-feynman}
\usepackage[section]{placeins}
\tikzfeynmanset{compat=1.0.0}

\definecolor{darkred}{rgb}{0.6,0,0}
\definecolor{darkpurple}{rgb}{0.5,0,0.5}
\def\hc{\text{h.c.}}

\newcommand{\Q}{Q}
\newcommand{\Op}{\mathcal{O}}
\newcommand{\pkg}[1]{{\tt #1}\xspace}
\newcommand{\dsix}{\pkg{DsixTools}}
\newcommand{\dsixv}{\pkg{DsixTools 2.0}}
\newcommand{\mathe}{\pkg{Mathematica}}

\begin{document}

\vspace*{-2cm}
\begin{flushright}
IFIC/19-60 \\
\vspace*{2mm}
\end{flushright}

\begin{center}
\vspace*{15mm}

\vspace{1cm}
{\Large \bf 
High-energy constraints from low-energy neutrino non-standard interactions
} \\
\vspace{1cm}

{\bf Jorge Terol-Calvo$^{\text{a}}$, Mariam T\'ortola$^{\text{a,b}}$, Avelino Vicente$^{\text{a}}$}

 \vspace*{.5cm} 
$^{\text{a}}$Instituto de F\'{\i}sica Corpuscular (CSIC-Universitat de Val\`{e}ncia), \\
C/ Catedr\'atico Jos\'e Beltr\'an 2, E-46980 Paterna (Val\`{e}ncia), Spain

 \vspace*{.3cm} 
$^{\text{b}}$Departament de F\'{\i}sica Te\`{o}rica, Universitat de Val\`{e}ncia, 46100 Burjassot, Spain

 \vspace*{.3cm} 
\href{mailto:jorge.terol@ific.uv.es}{jorge.terol@ific.uv.es}, \href{mailto:mariam@ific.uv.es}{mariam@ific.uv.es}, \href{mailto:avelino.vicente@ific.uv.es}{avelino.vicente@ific.uv.es}
\end{center}

\vspace*{10mm}
\begin{abstract}\noindent\normalsize
Many scenarios of new physics predict the existence of neutrino
Non-Standard Interactions, new vector contact interactions between
neutrinos and first generation fermions beyond the Standard Model. We
obtain model-independent constraints on the Standard Model Effective
Field Theory at high energies from bounds on neutrino non-standard
interactions derived at low energies. Our analysis explores a large
set of new physics scenarios and includes full one-loop running
effects below and above the electroweak scale. Our results show that
neutrino non-standard interactions already push the scale of new
physics beyond the TeV. We also conclude that bounds derived by other
experimental probes, in particular by low-energy precision
measurements and by charged lepton flavor violation searches, are
generally more stringent. Our study constitutes a first step towards
the systematization of phenomenological analyses to evaluate the
impact of neutrino Non-Standard Interactions for new physics scenarios
at high energies.
\end{abstract}




\section{Introduction} \label{sec:intro}

The Standard Model (SM) provides a successful description of a vast
amount of particle physics phenomena. This includes many predictions
at high-energy colliders, with the recent discovery of the Higgs boson
as the latest example, as well as an astonishing agreement with a long
list of precision measurements performed at low-energy
experiments. However, despite its success, there are several
well-known problems the SM cannot address. Among them, the existence
of non-zero neutrino masses is arguably the most robust one. After the
discovery of neutrino flavor oscillations, it has become clear that
the leptonic sector of the SM must be extended with some additional
states responsible for the generation of neutrino masses. Even though
the underlying physics is not known, this fact has been completely
established due to the high precision achieved in the determination of
the neutrino oscillation parameters~\cite{deSalas:2017kay}.

Many neutrino mass models have been proposed over the years. In most
scenarios, the new Beyond the Standard Model (BSM) degrees of freedom
have masses well above the electroweak scale, making them unreachable
to current colliders. In this case, one is allowed to use Effective
Field Theory (EFT) techniques, integrate out the heavy states and
describe their impact at low energies by means of a collection of
effective operators with canonical dimension larger than four. Following this procedure, in
addition to the well-known dimension-5 Weinberg operator that induces
Majorana neutrinos, one usually obtains other non-renormalizable
operators with potentially observable effects in low-energy
experiments.

In this work, we use
bounds on the so-called neutrino Non-Standard Interactions
(NSI)~\cite{Wolfenstein:1977ue,Valle:1987gv,Roulet:1991sm,Guzzo:1991hi,Farzan:2017xzy} derived at low-energy experiments to
set constraints valid at high energies. In order to do that, we make
use of the Standard Model Effective Field Theory
(SMEFT)~\cite{Buchmuller:1985jz,Grzadkowski:2010es} and the Low-Energy
Effective Field Theory (LEFT)~\cite{Jenkins:2017jig}, two well-known
EFTs valid at energies above or below the electroweak scale,
respectively. The link between neutrino NSI and well-established
EFTs, such as the LEFT and the SMEFT, allows one to study the
phenomenology of a wide class of New Physics (NP) scenarios in a
model-independent way and easily confront results coming from a large
diversity of experiments. In fact, this approach has been adopted in
many recent works, deriving bounds from low-energy scattering
\cite{Altmannshofer:2018xyo} or reactor \cite{Falkowski:2019xoe}
experiments, or studying the interplay with searches for lepton flavor
violating processes \cite{Davidson:2019iqh}. Lepton number violating
NSI have been considered in the context of the SMEFT in
\cite{Bolton:2019wta}, whereas a remarkable effort to provide a
consistent EFT description of NSI has been made in
\cite{Falkowski:2019xoe,Falkowski:2019kfn}. The generalization of
neutrino NSI including scalar or tensor couplings, the so-called
Neutrino Generalized Interactions (NGI)
\cite{Lindner:2016wff,AristizabalSierra:2018eqm}, have also been
discussed using an EFT language in \cite{Bischer:2019ttk}.

Embedding NSI (or NGI) into EFTs as well established as the SMEFT
and the LEFT provides a robust theoretical background and creates a
direct link to other phenomenological directions. Among other
advantages of this approach, one can easily compare the bounds
obtained from neutrino NSI to bounds derived by other experimental
probes. Here we will be interested in low-energy precision
measurements and charged lepton flavor violation. We will
systematically study a substantial region of the SMEFT parameter space
with the help of \dsix \cite{Celis:2017hod,dsixv}, a \mathe package
for the Renormalization Group Equations (RGE) running and matching in
the SMEFT and the LEFT. This tool allows us to include full one-loop
running effects (in the SMEFT and in the LEFT) in our numerical
analysis. As a result of this, we will obtain robust numerical results
and assess the relevance of RGE running for the NP scenarios
considered in our study.

The rest of the manuscript is structured as follows. In
Sec.~\ref{sec:NSIs} we review the formalism and current experimental
bounds on neutrino NSI. Sec.~\ref{sec:EFTs} introduces the SMEFT and
the LEFT, the two EFTs considered in our analysis, while
Sec.~\ref{sec:NSIsEFTs} shows how these theories can be used in
connection to neutrino NSI. Finally, we present our results in
Sec.~\ref{sec:results} and  conclude in
Sec.~\ref{sec:conclusions}. Additional definitions are given in
Appendix~\ref{sec:app1}.

\section{Neutrino non-standard interactions}
\label{sec:NSIs}
As commented in the introduction, new neutrino vector interactions beyond the Standard Model can arise from neutrino mass models and other  BSM theories.  In the low energy regime, neutrino NSI with matter fields can be formulated in terms of  an effective four-fermion Lagrangian as follows
\begin{align}
\mathcal{L}_{CC}^{NSI}=&-\dfrac{G_F}{\sqrt{2}}\left(\epsilon^{ff^\prime L}_{pr} \left[\bar{\nu}_p\gamma^\mu(1-\gamma_5)e_r\right]\left[\bar{f}\gamma_\mu(1-\gamma_5)f^\prime\right]\right.\nonumber\\&+\left. \epsilon^{ff^\prime R}_{pr} \left[\bar{\nu}_p\gamma^\mu(1-\gamma_5)e_r\right]\left[\bar{f}\gamma_\mu(1+\gamma_5)f^\prime\right]\right)\label{eq:CCNSI}
\end{align}
and
\begin{align}
\mathcal{L}_{NC}^{NSI}=&-\dfrac{G_F}{\sqrt{2}}\left(\epsilon^{f L}_{pr} \left[\bar{\nu}_p\gamma^\mu(1-\gamma_5)\nu_r\right]\left[\bar{f}\gamma_\mu(1-\gamma_5)f\right]\nonumber\right.\\&+\left.\epsilon^{f R}_{pr} \left[\bar{\nu}_p\gamma^\mu(1-\gamma_5)\nu_p\right]\left[\bar{f}\gamma_\mu(1+\gamma_5)f\right]\right), \label{eq:NCNSI}
\end{align}	
where $\epsilon^{ff^\prime L,R}_{pr}$ and $\epsilon^{f L,R}_{pr}$ are dimensionless coefficients that quantify the strength of the NSI between neutrinos of flavor $p$ and $r$ and the matter field $f,f^\prime=u,d$ with $f\neq f^\prime$ for the case of charged current (CC)-NSI and $f=e,u,d$ for neutral current (NC)-NSI.

Neutrino NSI can affect experiments at the neutrino production via CC-NSI, changing the flavor distribution of the initial neutrino flux, and detection via both CC and NC NSI, depending on the detection technique of the experiment. Besides, neutrino NC-NSI can affect their propagation through matter as well, modifying the effective matter potential felt by neutrinos. In this work, we will concentrate on NC-NSI.~\footnote{The effect of CC-NSI on reactor and long-baseline neutrino experiments has been discussed, for instance, in Refs.~\cite{Kopp:2007ne,  Agarwalla:2014bsa,Girardi:2014kca}. For a detailed analysis of CC-NSI in the context of EFTs, see \cite{Falkowski:2019kfn}.}

The potential signal of NSI on neutrino experiments has been analyzed in detail in the literature recently~\cite{Dev:2019anc,Farzan:2017xzy,Miranda:2015dra,Ohlsson:2012kf}. The impact of this signal on the extraction of neutrino oscillation parameters from experimental data has also been extensively discussed; see, for instance Refs.~\cite{Miranda:2004nb,Escrihuela:2009up, Coloma:2015kiu,Masud:2016gcl,Esteban:2018ppq}. However, since no signal of NSI has been experimentally reported yet, at the moment we only have upper bounds on their magnitude.
These limits come from a variety of neutrino experiments, from oscillation experiments using solar, atmospheric, reactor or accelerator neutrino sources, to laboratory experiments measuring  neutrino-electron and (coherent) neutrino-nucleus scattering. The size of the constraints on the NSI couplings depends on the neutrino flavors implied in the process, the most stringent one corresponding to the $\epsilon_{\mu\tau}^d$ coupling, bounded to be below 1\% (at 90\% C.L.) by the neutrino telescope IceCube~\cite{Aartsen:2017xtt}.

From the point of view of particle physics models, NSI are mainly thought to come from interactions of 
an ultraviolet (UV) complete theory mediated by a heavy particle $X$ of mass $m_X\gg m_{\rm EW}$. Other 
alternative approaches to generate these new interactions have also been proposed.  In particular, a possible
 explanation is to take the mass of the mediator particle much below the electroweak scale, 
 $m_X\ll m_{\rm EW}$~\cite{Farzan:2015doa,Farzan:2015hkd,Farzan:2016wym}. This choice can avoid the strong
  bounds coming from charged lepton processes, linked to NSI due to gauge invariance \cite{Gavela:2008ra}. This would allow the prediction of larger sizes of the NSI couplings, accessible to current or near future experiments.
In any case, here we will focus on the first possibility, where the EFT approach applies.

For completeness, we would like to comment on the possibility of
neutrino scalar and tensor four-fermion interactions, considered
lately in the
literature~\cite{Lindner:2016wff,AristizabalSierra:2018eqm,Bischer:2019ttk,Bischer:2018zcz}. Note,
however, that these NGI involve right-handed neutrinos and, therefore,
they are not relevant to our study since, as we will discuss in the
following section, our EFT analysis does not involve new particles.

\section{EFTs}
\label{sec:EFTs}

The SM of particle physics successfully describes a wide range of
phenomena. However, it still leaves some questions unanswered,
including the identity of the particle(s) accounting for dark matter
or how neutrino masses are generated. For this reason, it is common to
think of the SM as an effective theory valid up to a certain high-energy
scale $\Lambda_{\rm UV}$, at which some unknown NP degrees of freedom would
lie. The gauge group of a field theory valid above the electroweak
scale should contain the SM gauge group and full particle content and,
at energies below $\Lambda_{\rm UV}$, reduce to the SM. Generally, in
most theories beyond the SM, this reduction occurs via decoupling of
heavy particles with masses of order $\Lambda_{\rm UV}$ or
larger. This leads to the appearance of higher dimensional operators
in the SM Lagrangian suppressed by powers of $\Lambda_{\rm UV}$. The
EFT built with all the operators that respect the SM gauge group with
dimension $d \geq 5$ is the Standard Model Effective Field Theory
(SMEFT). The SMEFT Lagrangian is given by
\begin{equation}
  \mathcal{L}_{\mathrm{SMEFT}}=\mathcal{L}_{\mathrm{SM}}+\sum_{d \geq 5} \mathcal{L}^{(d)}_{\mathrm{SMEFT}},
\end{equation}	
with
\begin{equation}
  \mathcal{L}^{(d)}_{\mathrm{SMEFT}} = \sum_{i=1}^{n_d} \frac{C_i^{(d)}}{\Lambda_{\rm UV}^{d-4}} \, \Q_i^{(d)} \, ,
\end{equation}	
where $\mathcal{L}_{\mathrm{SM}}$ is the SM Lagrangian, $\Q_i^{(d)}$
are operators of dimension $d$ and $C_i^{(d)}$ the corresponding
Wilson coefficients (WCs). We note that the $C_i^{(d)}$ WCs have been
defined as dimensionless quantities by making explicit the suppression
by the high-energy scale $\Lambda_{\rm UV}$. The full set of SMEFT
operators up to dimension six was given in \cite{Grzadkowski:2010es},
defining the so-called \textit{Warsaw basis}. Finally, the complete
one-loop anomalous dimension matrix for the dimension-six operators in
this basis was obtained in
\cite{Jenkins:2013zja,Jenkins:2013wua,Alonso:2013hga,Alonso:2014zka}. This
describes the energy evolution of the Wilson coefficients as
\begin{equation}
  \mu\dfrac{dC_i}{d\mu}=\dfrac{1}{16\pi^2}\sum_j\gamma^{S}_{ij}C_j \, ,
\end{equation}
where $\mu$ is the renormalization scale and $\gamma^{S}$ the
anomalous dimension matrix for the operators of the SMEFT. Among all
the operators of the SMEFT, we list here the most relevant operators
for the study of neutrino NSI ($p,r,s,t$ are flavor indices):
\begin{multicols}{2}
  \begin{itemize}
  \item $\Q_{\substack{\ell \ell\\prst}}=\left[\bar{\ell}_{p}\gamma^\mu\ell_{r}\right]\left[\bar{\ell}_{s}\gamma_\mu\ell_{t}\right]$
  \item $\Q_{\substack{\ell e\\prst}}=\left[\bar{\ell}_{p}\gamma^\mu\ell_{r}\right]\left[\bar{e}_{s}\gamma_\mu e_{t}\right]$
  \item $\Q_{\substack{\ell q\\prst}}^{(1)}=\left[\bar{\ell}_{p}\gamma^\mu\ell_{r}\right] \left[\bar{q}_{s}\gamma_\mu q_{t}\right]$
  \item $\Q_{\substack{\ell u\\prst}}=\left[\bar{\ell}_{p}\gamma^\mu\ell_{r}\right]\left[\bar{u}_{s}\gamma_\mu u_{t}\right]$
  \item $\Q_{\substack{\ell q\\prst}}^{(3)}=\left[\bar{\ell}_{p}\gamma^\mu\tau^I\ell_{r}\right] \left[\bar{q}_{s}\gamma_\mu\tau^I q_{t}\right]$
  \item $\Q_{\substack{\ell d\\prst}}=\left[\bar{\ell}_{p}\gamma^\mu\ell_{r}\right]\left[\bar{d}_{s}\gamma_\mu d_{t}\right]$\
  \end{itemize}
\end{multicols}
\noindent where $\ell$ and $q$ are the SM lepton and quark doublets, and $e$, $u$ and $d$ the singlets.\newline

Neutrino experiments mainly deal with energies way below the
electroweak scale. Therefore, a new effective theory is needed to
describe low-energy processes. This EFT can be derived from the SM by
integrating out the massive electroweak gauge bosons ($W^{\pm},Z$),
the Higgs boson and the chiral top quark fermion fields ($t_L$ and
$t_R$). The gauge group of this Low-Energy Effective Field Theory
(LEFT) is $\rm SU(3)_C\times U(1)_Q$, i.e. the symmetry of QCD and
QED. The LEFT Lagrangian reads
\begin{equation}\label{eq:LEFT}
  \mathcal{L}_{\mathrm{LEFT}}= \mathcal{L}_{\mathrm{QCD+QED}}+ \mathcal{L}_{\cancel{L}}^{(3)}+\sum_{d \geq 5}\mathcal{L}^{(d)}_{\mathrm{LEFT}} \, .
\end{equation}	
The first term contains the QCD gauge interaction for 2 families of up
quarks and 3 of down quarks, the QED gauge interaction for these
quarks and the three charged lepton families, and their Dirac mass
terms,
\begin{align}
  \mathcal{L}_{\mathrm{QCD+QED}}=&-\dfrac{1}{4}G_{\mu\nu}^aG^{a,\mu\nu}-\dfrac{1}{4}F_{\mu\nu}F^{\mu\nu}+\theta_s\dfrac{g_s^2}{32\pi^2}G_{\mu\nu}^a\tilde{G}^{a,\mu\nu}+\theta_{\mathrm{QED}}\dfrac{e^2}{32\pi^2}F_{\mu\nu}\tilde{F}^{\mu\nu}+\nonumber\\
  &+\sum_{\psi=e,\nu_L,u,d} i\bar{\psi}\gamma^\mu D_\mu\psi-\sum_{\psi=e,u,d} \bar{\psi}_{Rr}\left[M_\psi\right]_{rs}\psi_{Ls}-\hc \, ,
\end{align}	
where $r,s$ are flavor indices and $D_\mu\psi=\left(\partial_\mu - i Q
A_\mu-i g_sT_s^a G^a_\mu\right)$. We note that the three left-handed
neutrinos are gauge singlets with no Dirac mass term. The second term
in Eq.~\eqref{eq:LEFT} consists of $\Delta L=\pm2$ Majorana mass terms
for the left-handed neutrinos,
\begin{equation}
  \mathcal{L}_{\cancel{L}}^{(3)}=-\dfrac{1}{2}\left[M_\nu\right]_{rs}\left(\nu_{Lr}^TC\nu_{Ls}\right)+\hc \, 
\end{equation}	
Here $M^T_\nu=M_\nu$ is the symmetric Majorana mass matrix. In the
case of 3 flavors of neutrinos, there will be 6 different $\Delta L=2$
operators and 6 conjugate $\Delta L=-2$ operators. Finally, the last
piece contains operators of dimension five or higher
\begin{equation}
  \mathcal{L}^{(d)}_{\mathrm{LEFT}} = \sum_{i=1}^{n_d} \frac{L_i^{(d)}}{v^{d-4}} \, \Op_i^{(d)} \, ,
\end{equation}	
where $\Op_i^{(d)}$ are dimension $d$ operators and $L_i^{(d)}$ their WCs
coefficients. Again, the $L_i^{(d)}$ WCs have been defined as
dimensionless quantities by introducing an explicit suppression by
$1/v^{d-4}$, where $v$ is the Higgs vacuum expectation value that sets
the electroweak scale. The full set of LEFT operators up to
dimension-6 and their tree-level matching relations with the SMEFT
operators can be found in \cite{Jenkins:2017jig}.~\footnote{We note
  that the one-loop SMEFT-LEFT matching relations were recently
  derived in \cite{Dekens:2019ept}, although we will not use them in
  our analysis.} We will stick to this basis of operators, referred to
as the \textit{San Diego basis}. The complete one-loop anomalous
dimension matrix for this basis of LEFT operators, $\gamma^{L}$, was
derived in \cite{Jenkins:2017dyc}, such that
\begin{equation}
  \mu\dfrac{dL_i}{d\mu}=\dfrac{1}{16\pi^2}\sum_j\gamma^{L}_{ij}L_j \, ,
\end{equation}
describes the evolution of the $L_i$ WCs with the renormalization
scale $\mu$. We now list some LEFT operators of relevance for the
study of neutrino NC NSI: 
\begin{multicols}{2}
  \begin{itemize}
  \item $\Op^{V,LL}_{\substack{\nu u \\prst}}= \left[\bar{\nu}_{L,p}\gamma^\mu\nu_{L,r}\right]\left[\bar{u}_{L,s}\gamma_\mu u_{L,t}\right]$  
  \item $\Op^{V,LL}_{\substack{\nu d\\prst}}= \left[\bar{\nu}_{L,p}\gamma^\mu\nu_{L,r}\right]\left[\bar{d}_{L,s}\gamma_\mu d_{L,t}\right]\quad$  
  \item $\Op^{V,LL}_{\substack{\nu e\\prst}}=\left[\bar{\nu}_{L,p}\gamma^\mu\nu_{L,r}\right]\left[\bar{e}_{L,s}\gamma_\mu e_{L,t}\right]$  
  \item $\Op^{V,LR}_{\substack{\nu u \\prst}}= \left[\bar{\nu}_{L,p}\gamma^\mu\nu_{L,r}\right]\left[\bar{u}_{R,s}\gamma_\mu u_{R,t}\right]$
  \item $\Op_{\substack{\nu d\\prst}}^{V,LR}= \left[\bar{\nu}_{L,p}\gamma^\mu \nu_{L,r}\right]\left[\bar{d}_{R,s}\gamma_\mu d_{R,t}\right]$
  \item $\Op^{V,LR}_{\substack{\nu e\\prst}}=\left[\bar{\nu}_{L,p}\gamma^\mu\nu_{L,r}\right]\left[\bar{e}_{R,s}\gamma_\mu e_{R,t}\right]$    
  \end{itemize}
\end{multicols}
\noindent where  $\nu_L$, $e_{L/R}$, $u_{L/R}$ and $d_{L/R}$ are the chiral left/right-handed neutrino, charged lepton, up quark and down quark fields. \newline


The electroweak scale sets the limit between the SMEFT and the LEFT
and determines the energy scale at which these two theories must be
matched by integrating out the $W$ and $Z$ gauge bosons, the Higgs boson and the top quark. When doing so, one must take into account the breaking of the
electroweak symmetry, therefore matching the SMEFT in the \emph{broken
  phase} with the LEFT.~\footnote{Appendix~\ref{sec:app1} compiles the
  most relevant analytical expressions for the SM parameters including
  their modifications in the presence of contributions from
  dimension-6 SMEFT operators.} An obvious feature arising from this
matching will be the breaking of the $\rm SU(2)_L$ doublets,
originating several LEFT operators from a single SMEFT operator. For
example, from the SMEFT operator $\Q_{\ell \ell}$, the LEFT operators
$\Op^{V,LL}_{\nu e}$, $\Op^{V,LL}_{e e}$ and $\Op^{V,LL}_{\nu \nu}$
will emerge. Furthermore, the LEFT operators will receive several
contributions. In addition to those originated from the dimension-6
SMEFT operators, pure SM contributions exist as well. For instance,
the matching relation for the $\Op^{V,LL}_{\nu \nu}$ LEFT operator is
\begin{equation} \label{eq:matchingexample}
  \dfrac{L^{V,LL}_{\substack{\nu \nu\\prst}}}{v^2}=\dfrac{C_{\substack{\ell \ell\\prst}}}{ \Lambda_{\rm UV}^2}-\dfrac{\bar{g}_Z^2}{4 M_Z^2}\left[Z_\nu\right]_{pr}\left[Z_\nu\right]_{st}-\dfrac{\bar{g}_Z^2}{4 M_Z^2}\left[Z_\nu\right]_{pt}\left[Z_\nu\right]_{sr} \, ,
\end{equation}
where $p,r,s,t$ are flavor indices. The first term constitutes the
contribution of the SMEFT operator $\Q_{\ell \ell}$, whereas the last
two terms correspond to two contributions to the $\Op^{V,LL}_{\nu
  \nu}$ LEFT operator obtained by $Z$ boson
exchange. $\left[Z_\nu\right]$ is the $Z$ coupling to a pair of
neutrinos which, in addition to the pure SM coupling, contains
contributions from the SMEFT operators $\Q^{(1)}_{H \ell}$ and
$\Q^{(3)}_{H \ell}$. Finally, $\bar{g}_Z$ is an effective coupling
containing the contribution of dimension-6 Higgs-gauge-boson operators
$X^2H^2$. Eq.~\eqref{eq:matchingexample} assumes the SMEFT WCs to be
given in the fermion \textit{Up basis}, defined by diagonal up-quark
and charged lepton Yukawa matrices, since this basis allows one to
identify the top quark, one of the fields integrated out at this
stage. We will adopt this implicit assumption in all the matching
relations given in this paper and omit the unitary matrices that
transform to the Up basis in order to simplify the resulting
expressions. The full set of SMEFT-LEFT tree-level matching relations
can be found in \cite{Jenkins:2017jig}.~\footnote{At energies below
  $\mu \sim 5$ GeV one should adopt other EFTs, better suited to take
  into account the non-perturbative nature of the strong interactions
  in this energy regime (see for instance
  \cite{Dekens:2018pbu,Falkowski:2019xoe}). In order to simplify our
  study and be able to ignore this issue, we will never run below $5$
  GeV and neglect this possibility.}
  \section{Neutrino NSI in the LEFT and the SMEFT}
  \label{sec:NSIsEFTs}
  
  After discussing neutrino NSI and two EFTs of interest, the SMEFT and
  the LEFT, we proceed to establish a link between them. This will allow
  us to study neutrino NSI in the language of the SMEFT and the LEFT
  and, more importantly, to make use of the theoretical machinery
  developed for these two theories. In fact, as shown in
  Sec.~\ref{sec:NSIs}, neutrino NSI are encoded by a set of
  coefficients of low-energy effective operators. Therefore, the link to
  the LEFT is quite straightforward. One can find a one-to-one relation
  between the NSI effective operators and the LEFT operators which, in
  turn, can be matched to the SMEFT operators valid at high energies. In
  Table~\ref{ta:Matching}, we list all the NC NSI coefficients and their
  matching with the LEFT and SMEFT WCs. This table makes use of the
  definitions in Appendix~\ref{sec:app1}, which are taken from
  \cite{Jenkins:2017jig}. As explained in Sec.~\ref{sec:EFTs}, the SMEFT
  WCs are assumed to be given in the fermion Up basis. The relevant
  unitary matrices involved in the transformation to this basis are not
  explictly indicated to simplify the notation.~\footnote{Alternatively,
    \cite{Bischer:2019ttk} gives analogous matching relations in the
    Down basis, in which the down-quark Yukawa matrix is diagonal. This
    reference also includes explicitly the quark mixing matrices
    appearing in the transformation to the Down basis.} Furthermore, we
  note that the SMEFT-LEFT matching relations in Table~\ref{ta:Matching}
  include pure SM contributions. These must be removed in the final
  matching to the NSI coefficients to properly identify the non-standard
  pieces.
    
  {
  \renewcommand{\arraystretch}{2.0}
  \begin{table}[!t]
    \centering
    \begin{tabular}{ccc}
      \hline\hline
      {\bf NSI} & {\bf LEFT} & {\bf SMEFT} \\
      \hline\hline
      $-2\sqrt{2}G_F\, \epsilon^{u L}_{pr}$   &  $\dfrac{1}{v^2} \, L^{V,LL}_{\substack{\nu u\\pr11}}$  & $\dfrac{1}{\Lambda_{\rm UV}^2} \left(C^{(1)}_{\substack{\ell q\\pr11}}+C^{(3)}_{\substack{\ell q\\pr11}}\right)-\dfrac{\Bar{g}_Z^2}{M_Z^2}\left[Z_\nu\right]_{pr}\left[Z_{u_L}\right]_{11}$ \vspace*{0.2cm} \\ 		
      $-2\sqrt{2}G_F\, \epsilon^{d L}_{pr}$   &  $\dfrac{1}{v^2} \, L^{V,LL}_{\substack{\nu d\\pr11}}$  &  $\dfrac{1}{\Lambda_{\rm UV}^2} \left(C^{(1)}_{\substack{\ell q\\pr11}}-C^{(3)}_{\substack{\ell q\\pr11}}\right)-\dfrac{\Bar{g}_Z^2}{M_Z^2}\left[Z_\nu\right]_{pr}\left[Z_{d_L}\right]_{11}$ \vspace*{0.2cm} \\ 
      $-2\sqrt{2}G_F\, \epsilon^{e L}_{pr}$   &  $\dfrac{1}{v^2} \, L^{V,LL}_{\substack{\nu e\\pr11}}$  &  $\dfrac{1}{\Lambda_{\rm UV}^2} \left(C_{\substack{\ell \ell\\pr11}}+C_{\substack{\ell \ell\\ 11pr}}\right)-\dfrac{\Bar{g}_2^2}{2 M_W^2}\left[W_\ell\right]_{p1}\left[W_{\ell}\right]^*_{r1}-\dfrac{\Bar{g}_Z^2}{M_Z^2}\left[Z_\nu\right]_{pr}\left[Z_{e_L}\right]_{11}$ \vspace*{0.2cm} \\
      $-2\sqrt{2}G_F\, \epsilon^{u R}_{pr}$  &  $\dfrac{1}{v^2} \, L^{V,LR}_{\substack{\nu u\\pr11}}$  &  $\dfrac{1}{\Lambda_{\rm UV}^2} \, C_{\substack{\ell u\\pr11}}-\dfrac{\Bar{g}_Z^2}{M_Z^2}\left[Z_\nu\right]_{pr}\left[Z_{u_R}\right]_{11}$ \vspace*{0.2cm} \\    
      $-2\sqrt{2}G_F\, \epsilon^{d R}_{pr}$  &  $\dfrac{1}{v^2} \, L^{V,LR}_{\substack{\nu d\\pr11}}$  &  $\dfrac{1}{\Lambda_{\rm UV}^2} \, C_{\substack{\ell d\\pr11}}-\dfrac{\Bar{g}_Z^2}{M_Z^2}\left[Z_\nu\right]_{pr}\left[Z_{d_R}\right]_{11}$ \vspace*{0.2cm} \\    
      $-2\sqrt{2}G_F\, \epsilon^{e R}_{pr}$   &  $\dfrac{1}{v^2} \, L^{V,LR}_{\substack{\nu e\\pr11}}$  &  $\dfrac{1}{\Lambda_{\rm UV}^2} \, C_{\substack{\ell e\\pr11}}-\dfrac{\Bar{g}_Z^2}{M_Z^2}\left[Z_\nu\right]_{pr}\left[Z_{e_R}\right]_{11}$ \vspace*{0.2cm} \\
  \hline						
    \end{tabular}
    \caption{Tree-level matching of the NC NSI coefficients (with flavor indices $p$ and $r$) to the LEFT
      and SMEFT Wilson coefficients. The SMEFT-LEFT matching relations were derived in
      \cite{Jenkins:2017jig}. The SMEFT WCs are assumed to be given in
      the Up fermion basis, see Sec.~\ref{sec:EFTs} for details. The
      pure SM contributions are removed in the final matching to the NSI
      coefficients. We refer to Appendix~\ref{sec:app1} for notation and
      conventions.
      \label{ta:Matching}}
  \end{table}
  }
  
  Armed with these matching relations and the RGEs of both EFTs we can
  bring the bounds coming from low (high) energy experiments to high
  (low) energies and \emph{translate} them to the most convenient
  effective theory in each case. In particular, the main goal of our
  work is to use NC neutrino NSI to derive limits on the SMEFT WCs at
  high energies. One could naively think that the best
  method to do this is to start at low energies, match the NSI
  coefficients to the LEFT WCs, run up to the EW scale, match the LEFT
  to the SMEFT and finally run up to the high-energy scale $\Lambda_{\rm
    UV}$, where the resulting limit is obtained. However, this method is
  inconsistent because it does not take into account that the LEFT is
  more general than the SMEFT. For instance, a low-energy scenario with
  a single NSI coefficient or, equivalently, with a single LEFT WC, might
  not be consistent with a high-energy SMEFT origin, since $\rm SU(2)_L$
  gauge invariance imposes relations between LEFT WCs after SMEFT-LEFT
  matching. For this reason, running up from low energies is not (in
  general) a consistent approach to achieve our goal. Instead, one must
  do the opposite: to consider a SMEFT parameter point at high energies,
  run down to the EW scale, match to the LEFT, continue running down to
  low energies, match to the NSI coefficients and compare the resulting
  values to the experimental bounds. This process can be repeated for
  different input SMEFT scenarios and high-energy scales, thus
  determining the region of the SMEFT parameter space that is compatible
  with the low-energy NSI bounds. \\
    
  We will now illustrate our procedure with an explicit example.
    
  \subsection{An example}
    
  To make more explicit the method used in our analysis, we will now
  follow step by step the full path that takes from the high-energy
  SMEFT to the NSI coefficients that play a role in neutrino
  experiments. Let us study a process of interest such as a hypothetical
  new interaction between electron neutrinos and left-handed electrons.
    
\begin{figure}[t!]
    \centering
    \includegraphics[width=0.97\linewidth]{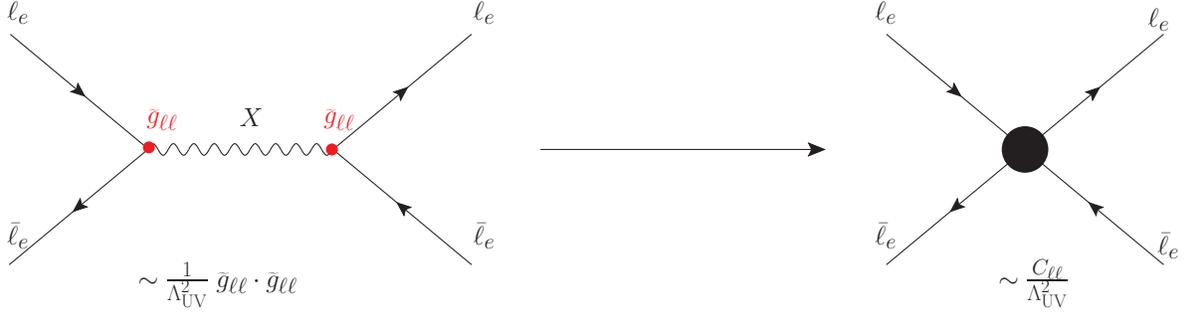}
    \caption{Feynman diagram of the process $\ell_e\bar{\ell}_e\to\ell_e\bar{\ell_e}$ via the exchange of an unknown heavy mediator that gives rise to a dimension-6 effective operator in the SMEFT.}
    \label{fig:diagram_NPtoSMEFT}
  \end{figure}

  First, we start at the high-energy scale $\Lambda_{\rm UV}$, where the
  $\rm SU(2)_L$ symmetry is unbroken, and consider an interaction
  involving the first generation lepton doublet, $\ell_e$, induced by
  the exchange of an unknown heavy vector mediator, $X$, with $m_X\sim
  \Lambda_{\rm UV}$. At energies below $m_X$, the tree-level exchange of
  the heavy $X$ vector can be effectively described by a 4-fermion
  interaction, with strength
  $\Tilde{g}_{\ell\ell}\cdot\Tilde{g}_{\ell\ell}$, where
  $\Tilde{g}_{\ell\ell}$ is the coupling of $X$ to a pair of lepton
  doublets, and suppressed by $\dfrac{1}{\Lambda_{\rm UV}^2}$. In Figure
  \ref{fig:diagram_NPtoSMEFT} we can see a diagrammatic representation
  of this.
  There, we see that integrating out $X$ leads to the generation of
  the 4-fermion SMEFT operator $Q_{\ell
    \ell}=\left[\Bar{\ell}_{p}\gamma^\mu\ell_{r}\right]\left[\Bar{\ell}_{s}\gamma_\mu\ell_{t}\right]$,
  with Wilson coefficient $C_{\ell
    \ell}=\Tilde{g}_{\ell\ell}\cdot\Tilde{g}_{\ell\ell}$ and suppressed
  by $1/\Lambda_{\rm UV}^2$. We can now solve the SMEFT RGEs to
  obtain the SMEFT Lagrangian at the electroweak scale, where the SM
  symmetry breaks and the SMEFT must be matched to the LEFT. With only
  one non-vanishing input SMEFT WC, the main contributions to the RGEs
  come from the terms proportional to it. For the case under discussion,
  the main term is
  \begin{equation}
    \dot{C}_{\substack{\ell \ell\\1111}}\sim 2 \, \left[Y_e^\dagger Y_e\right]_{11} \, C_{\substack{ \ell \ell\\ 1111}} \, .
  \end{equation}
  
\begin{figure}[t!]
    \centering
    \includegraphics[width=0.97\linewidth]{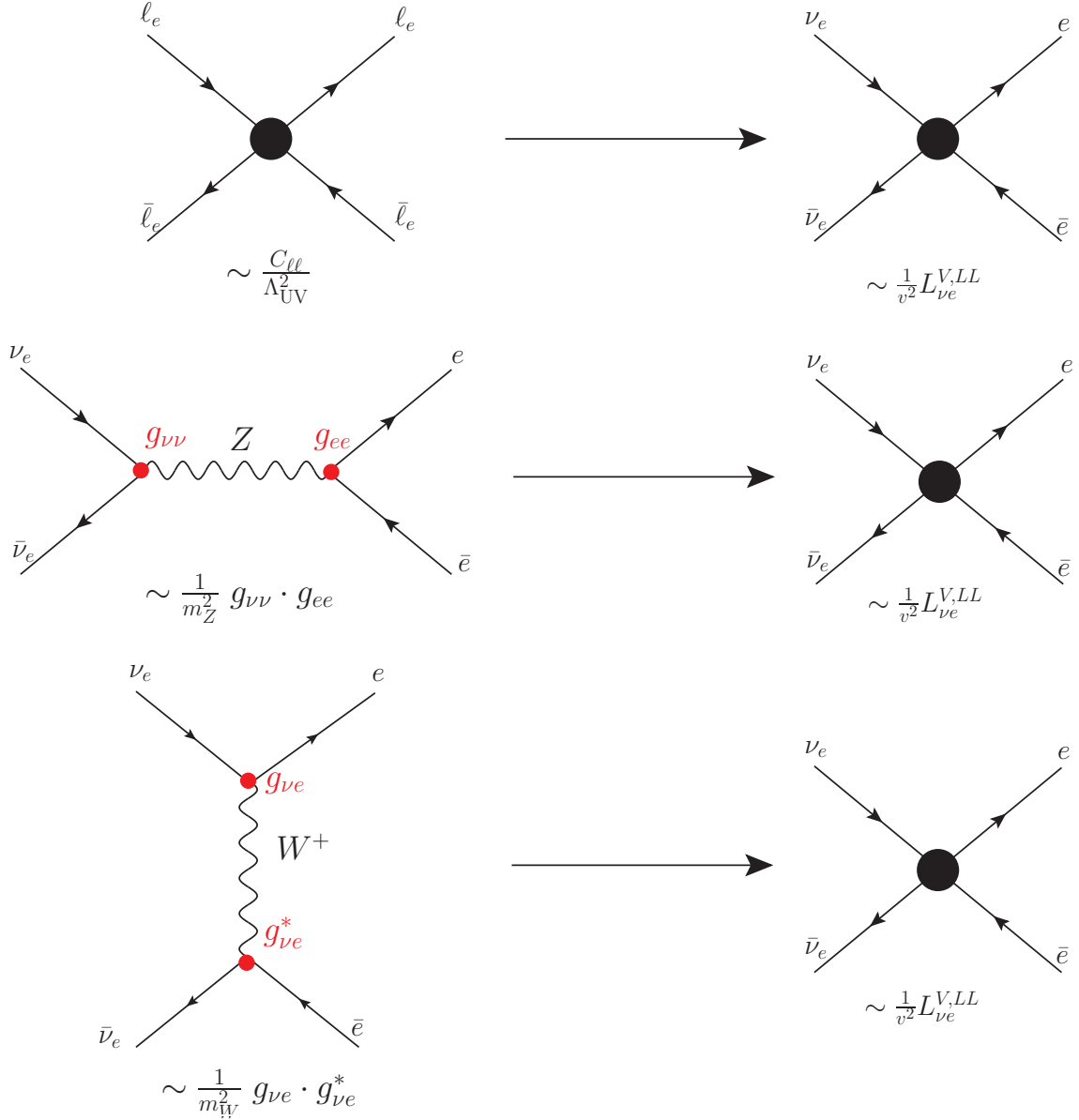}	
    \caption{Feynman diagrams that contribute to the process
      $\ell_e\bar{\ell}_e\to\ell_e\bar{\ell_e}$ in the SMEFT at
      tree-level, both from dimension-6 operators and from gauge boson
      exchange, giving rise to a dimension-6 effective operator in the
      LEFT.
      \label{fig:diagram_SMEFTtoLEFT}}
  \end{figure}

  The next step is to match the SMEFT with the LEFT. As already
  discussed, from a single SMEFT operator one gets several LEFT
  operators. In this case, since we are interested in electron
  neutrino-electron interactions, we focus on $\mathcal{O}^{V,LL}_{\nu
    e}$. We are then studying a 4-fermion diagram with two electron
  neutrinos and two left-handed electrons. An important feature to take
  into account is that this diagram is not only generated by the
  previously discussed dimension-6 SMEFT operator, but also from pure SM
  diagrams with $W$ and $Z$ bosons exchange, as we can see in Figure \ref{fig:diagram_SMEFTtoLEFT}. 
  The tree-level matching relation between this LEFT operator and the
  SMEFT is given by
  \begin{equation}
     \dfrac{L^{V,LL}_{\substack{\nu e\\ 1111}}}{v^2} = 2 \, \dfrac{C_{\substack{\ell \ell\\1111}}}{\Lambda_{\rm UV}^2} - \dfrac{\bar{g}_2^2}{2 M_W^2} \, \left[W_\ell\right]_{11} \, \left[W_{\ell}\right]^*_{11}-\dfrac{\bar{g}_Z^2}{M_Z^2} \, \left[Z_\nu\right]_{11} \, \left[Z_{e_L}\right]_{11} \, , 
  \end{equation}
  where $\left[W_{\ell}\right]$ and $\left[Z_{\nu/e_L}\right]$ are the
  $W$ and $Z$ couplings to neutrinos/electrons and $\bar{g}_2^2$ and
  $\bar{g}_Z^2$ quantify the strength of the $W$ and $Z$
  interactions. Explicit expressions for these couplings, including the
  corrections due to dimension-6 operators, are given in
  Appendix~\ref{sec:app1}.
  
  After the SMEFT-LEFT matching, we solve the RGEs down to the low-energy
  scale $\Lambda_{\rm IR}$, where the NSI bounds are set. For the
  operator we are interested in, the main RGE term is
  \begin{align}
  \dot{L}^{V,LL}_{\substack{\nu e\\1111}}\sim \dfrac{4}{3} \, e^2 \, L^{V,LL}_{\substack{ \nu e \\11 rr}} \, ,
  \end{align}
  where we sum over the flavor index $r$. Finally, we perform the
  matching between the LEFT and NSI operators. In this case, the relation
  between the coefficients is quite simple, as shown in
  Table~\ref{ta:Matching}. For the example we are going through, the
  matching is given by
  \begin{equation}\label{eq:Epsilonee}
  \epsilon^{e L}_{ee} = - \dfrac{\left(L^{V,LL}_{\substack{\nu e\\ 1111}}\right)^{\text{BSM}}}{2\sqrt{2} \, G_F \, v^2} \, ,
  \end{equation}
  where $\left(L^{V,LL}_{\substack{\nu e\\ 1111}}\right)^{\text{BSM}}$ is the pure BSM contribution to the LEFT WC $L^{V,LL}_{\substack{\nu e\\ 1111}}$. After this plain matching, we get a value for the NSI coefficient we
  are interested in at a certain energy scale, ready to be compared with
  the experimental bounds.
  
  \section{Numerical analysis}
\label{sec:results}

Having set our notation and described our strategy, we now proceed to
show the results of our numerical analysis. In order to explore a
substantial region of the huge SMEFT parameter space, we have
considered a large number of SMEFT operators (with specific flavor
indices) and applied the approach discussed in
Sec.~\ref{sec:NSIsEFTs}. This way, we have been able to \textit{map} a
region of interest in the SMEFT onto the NSI parameter space, where
the experimental constraints previously derived in the literature can
be directly applied.

A total of $112$ initial non-zero SMEFT WCs have been selected. For
each of them, and assuming only one at a time, we have considered $14$
different values for the NP scale, $\Lambda_{\rm UV}$, in the $[0.5,
  14]$ TeV range. This NP scale not only sets the starting point for
the RGE running, but also the value of the SMEFT WC, taken to be
precisely $|C_i| = 1$ at $\mu = \Lambda_{\rm UV}$. Then, as a result
of the strategy explained in Sec.~\ref{sec:NSIsEFTs}, $48$ NSI
coefficients are obtained at the low-energy scale $\Lambda_{\rm IR} =
5$ GeV for each scenario. This includes NC NSI with both
chiralities. We finally compare these values with the current NSI
experimental bounds and derive limits for the original SMEFT WCs at
high energies.

Our numerical calculations have been obtained with the help of \dsix
\cite{Celis:2017hod,dsixv}. This \mathe package has several tools and
functionalities for the RGE running and matching in the SMEFT and the
LEFT and is perfectly suited for our phenomenological exploration. In
particular, we used version 2.0~\cite{dsixv}, which fully integrates
the LEFT, and only added the matching between the LEFT and NSI
operators. This approach allows one to explore the relation between
neutrino NSI, the LEFT and the SMEFT in a systematic way. To the best
of our knowledge, our work is the first to study such connection
including full one-loop running effects. We use \dsix to solve the
LEFT RGEs numerically while the SMEFT RGEs are solved following a
semi-analytical approach based on an evolution matrix
formalism~\cite{Brivio:2019irc}. We take advantage of one of the main
\dsix functionalities: user-friendly input and output, which can be
given in the \dsix native format as well as using the {\tt WCxf}
exchange format~\cite{Aebischer:2017ugx}. All input parameters will be
assumed to be specified in the Up basis. We have explicitly checked
that the charged lepton Yukawa matrix remains in very good
approximation diagonal after RGE running, at high and low
energies. This allows for an easy identification of the neutrino
flavor eigenstates, precisely defined by the basis in which the
charged lepton Yukawas are diagonal. Furthermore, \dsix transforms all
SMEFT parameters to the Up basis before applying the matching
relations of Ref.~\cite{Jenkins:2017jig}.

In what concerns the experimental NSI limits used in our analysis,
these come from various sources, including neutrino oscillation and
scattering experiments.  We have used the bounds compiled in
\cite{Farzan:2017xzy}, where an extensive review of the NSI formalism
and experimental limits is done. More precisely, the most relevant
couplings for our analysis come from
\begin{itemize}
\item the analysis of neutrino-nucleon scattering data~\cite{Escrihuela:2011cf,Miranda:2015dra} ($\epsilon_{e\mu}^{qL}$).
\item the combined analysis of atmospheric and neutrino-nucleon scattering data~\cite{Escrihuela:2011cf} ($\epsilon_{\mu\mu}^{dV}$). 
\item the analysis of the atmospheric neutrino signal in IceCube DeepCore~\cite{Aartsen:2017xtt} ($\epsilon_{\mu\tau}^{qV}$).
\item the combined analysis of solar and KamLAND reactor data~\cite{Bolanos:2008km} ($\epsilon_{ee}^{eL}$). 
\item the combined analysis of reactor and accelerator data~\cite{Davidson:2003ha,Barranco:2007ej} ($\epsilon_{\mu\mu}^{eL}$). 
\item the combination of oscillation and coherent neutrino-nucleus scattering data~\cite{Coloma:2017ncl} ($\epsilon_{ee}^{uV}$). 
\item the analysis of atmospheric neutrino data~\cite{GonzalezGarcia:2011my,Farzan:2017xzy} ($\epsilon_{\tau\tau}^{qV}$). 
\end{itemize}

\subsection{Neutrino NSI from the SMEFT at high energies}
\label{subsec:NSIsResults}

Before moving to the discussion of the limits on SMEFT WCs derived
from neutrino NSI, it is illustrative to show some selected examples
of the NSI coefficients generated by several SMEFT scenarios. We can
visualize these results by plotting the values of the NSI coefficients
obtained for a certain initial non-zero SMEFT WC at several scales
along with the experimental bounds on the NSI coefficients. This way
we can easily determine whether a specific SMEFT scenario is
constrained or not due to neutrino NSI for a given $\Lambda_{\rm
  UV}$.
\begin{figure}[t]
  \centering
  \includegraphics[width=0.97\linewidth]{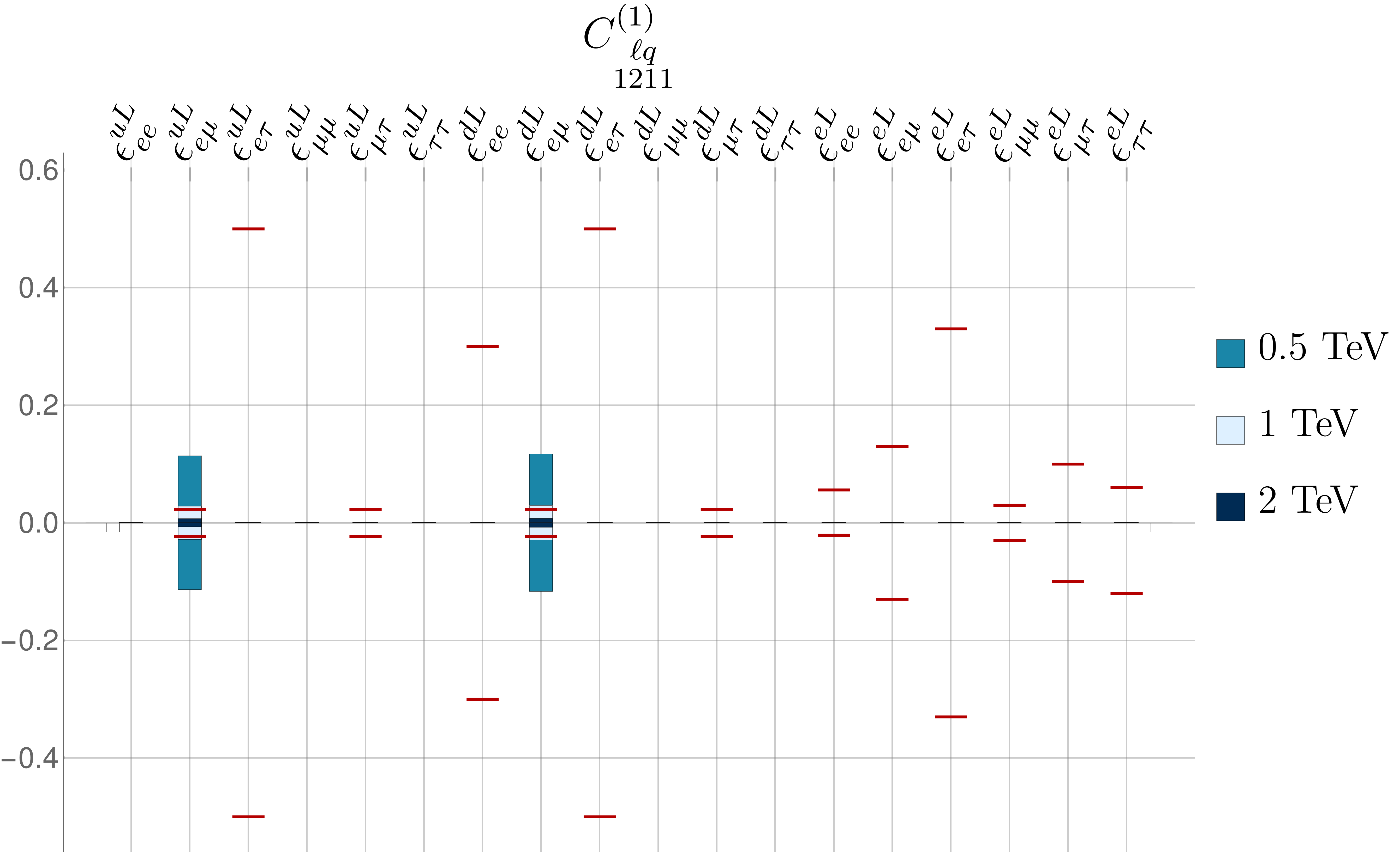}
  \caption{Values for the various chiral left NSI coefficients
    obtained assuming the SMEFT WC $| C^{(1)}_{\protect\substack{\ell q
        \\ 1211}} | = 1$ at $\mu = \Lambda_{\rm UV}$ for different
    values of $\Lambda_{\rm UV}$. The red lines correspond to the
    experimental bounds on the NSI coefficients compiled in
    \cite{Farzan:2017xzy}.}
  \label{fig:Plot1}
\end{figure}
A first example of this strategy is presented in Fig.~\ref{fig:Plot1}. This figure displays the
values of the left chiral NC NSI coefficients arising from the SMEFT
WC $C^{(1)}_{\substack{\ell q \\ 1211}}$. Three possible $\Lambda_{\rm
  UV}$ values are considered, $0.5$, $1$ and $2$ TeV. As expected,
lower NP scales imply larger NSI coefficients. The current
experimental limits on the different NSI coefficients are indicated
with red lines, implying the exclusion of any SMEFT parameter point
leading to NSI coefficients at low energies that fall outside of
them. For instance, in this example we find that the bound on
$\epsilon^{u L}_{e\mu}$ excludes $\Lambda_{\rm UV} \lesssim 1$
TeV. Also, this figure seems to indicate that only two NSI
coefficients are generated at $\Lambda_{\rm IR}$. Actually, since we
are including full one-loop running effects in our calculation, many
NSI coefficients are non-vanishing at low energies. For this
particular example, all $\mu-e$ flavor violating NSI coefficients are
generated. However, most of them are too small to be visualized in
Fig.~\ref{fig:Plot1} and only $\epsilon^{u L}_{e\mu}$ and $\epsilon^{d
  L}_{e\mu}$, the two $\mu-e$ flavor violating NSI coefficients with
first generation left-handed quarks, have sizable values. In fact, the
sizable values obtained for these two NSI coefficients could have been
\textit{predicted} just by using the tree-level matching relation
\begin{equation}
\epsilon^{u L}_{e\mu} \simeq \epsilon^{d L}_{e\mu} \sim - \dfrac{C^{(1)}_{\substack{\ell q \\ 1211}}}{2\sqrt{2} \, G_F \, \Lambda_{\rm UV}^2} \, ,
\end{equation}
given in Tab.~\ref{ta:Matching}. This approximate relation is
reproduced in our numerical results. Moreover, other NSI coefficients
would only be generated due to operator mixing effects. Since they
have tiny values, we conclude that operator mixing effects are
negligible in this scenario.

\begin{figure}[t]
  \centering
  \includegraphics[width=0.97\linewidth]{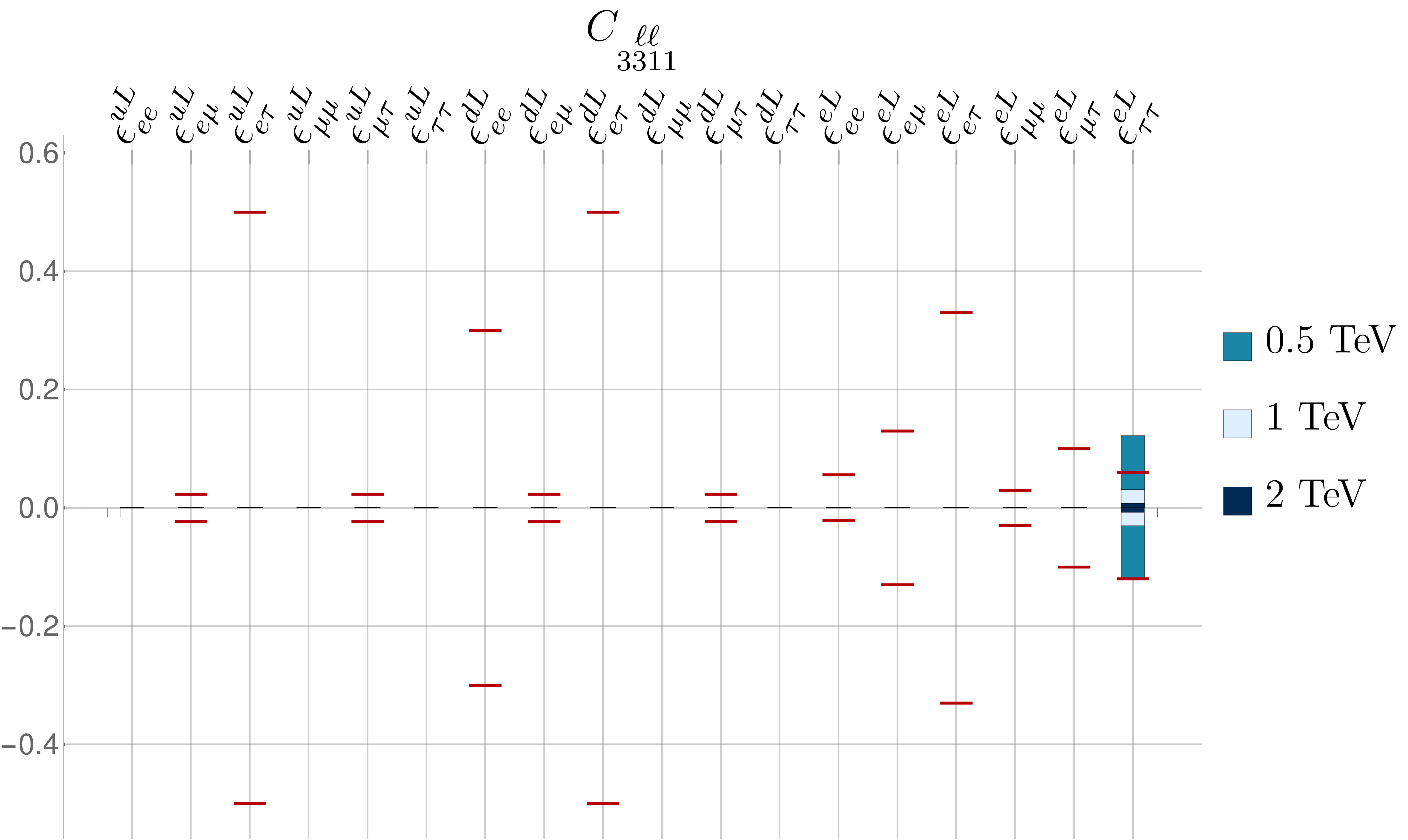}
  \caption{Values for the various chiral left NSI coefficients
    obtained assuming the SMEFT WC $|C_{\protect\substack{\ell \ell
        \\ 3311}} |= 1$ at $\mu = \Lambda_{\rm UV}$ for different
    values of $\Lambda_{\rm UV}$. The red lines correspond to the
    experimental bounds on the NSI coefficients compiled in
    \cite{Farzan:2017xzy}.}
  \label{fig:Plot2}
\end{figure}

\begin{figure}[t]
  \centering
  \includegraphics[width=0.97\linewidth]{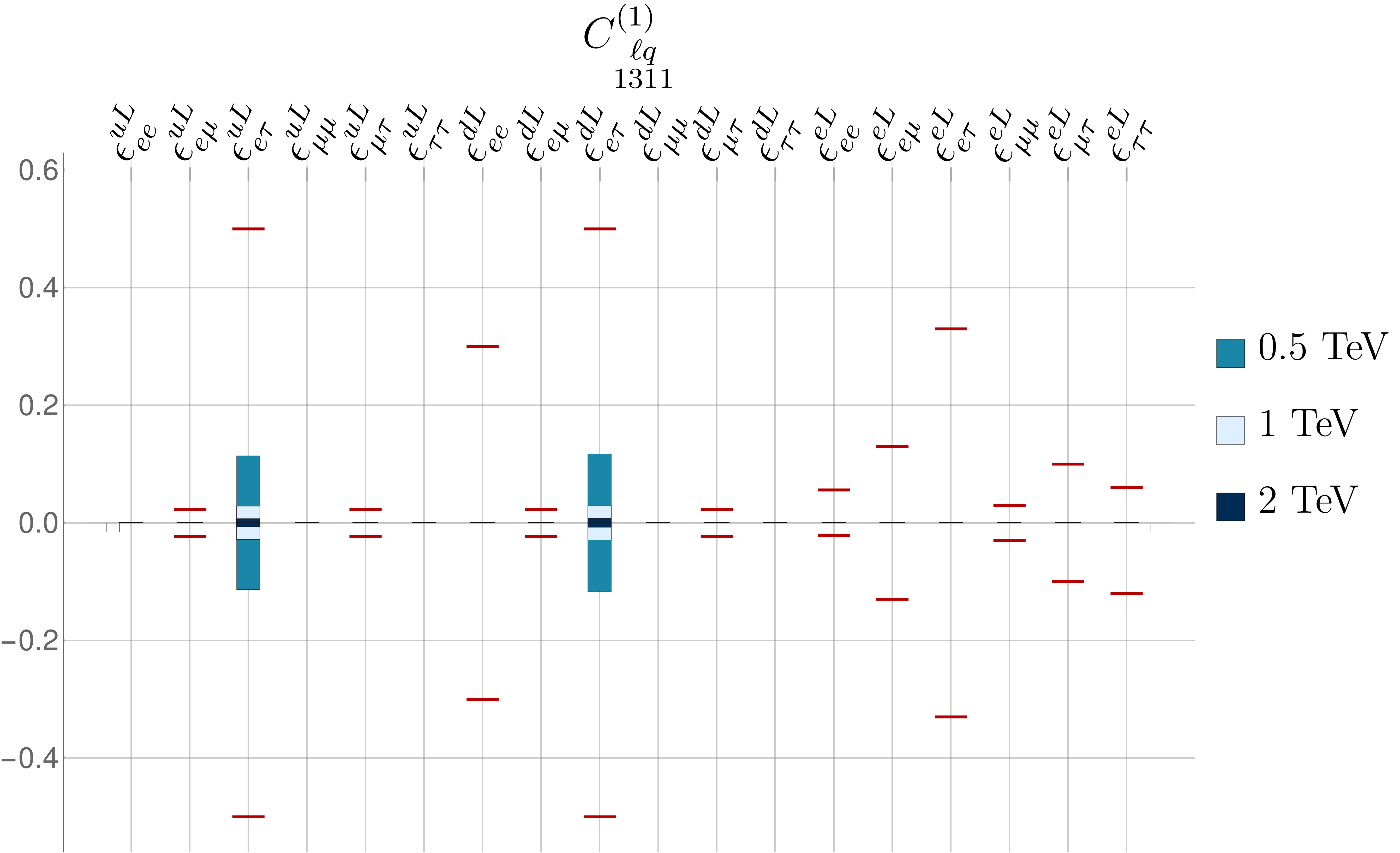}
  \caption{Values for the various chiral left NSI coefficients
    obtained assuming the SMEFT WC $|C^{(1)}_{\protect\substack{\ell q
        \\ 1311}}| = 1$ at $\mu = \Lambda_{\rm UV}$ for different
    values of $\Lambda_{\rm UV}$. The red lines correspond to the
    experimental bounds on the NSI coefficients compiled in
    \cite{Farzan:2017xzy}.}
  \label{fig:Plot5}
\end{figure}

One can also find SMEFT scenarios leading to NSI coefficients
compatible with the current limits even for NP scales as low as
$\Lambda_{\rm UV} = 0.5$ TeV.~\footnote{$\Lambda_{\rm UV} = 0.5$ TeV
  is the lowest NP scale considered in our analysis. Below that value
  the SMEFT approach is no longer justified.} This is the case when the
input SMEFT WC involves only 2nd or 3rd generation quarks. Even though
non-vanishing NSI coefficients with 1st generation quarks are obtained
due to quark mixing effects in the RGEs, these are always
tiny. Therefore, scenarios of this sort will not be considered in our
subsequent analysis, since they cannot be effectively bounded by
neutrino NSI. Similarly, there are scenarios leading to sizable NSI,
but not large enough to be constrained. This is for instance
illustrated in Fig.~\ref{fig:Plot2}, which shows the left chiral NC
NSI coefficients arising from the input SMEFT WC $C_{\substack{\ell
    \ell \\ 3311}}$. The only non-negligible NSI coefficient in this
case is $\epsilon^{e L}_{\tau\tau}$. The joint analysis of solar
neutrino experiments (mostly Super-Kamiokande) and KamLAND require
$-0.12 < \epsilon^{e L}_{\tau\tau} < 0.06$ at 90\% C.L. Due to the
asymmetry in these experimental limits, scenarios with
$C_{\substack{\ell \ell \\ 3311}} < 0$ require $\Lambda_{\rm UV}$ to
be above $\sim 0.7$ TeV, while the NP scale can be as low as $0.5$ TeV
when $C_{\substack{\ell \ell \\ 3311}} > 0$. Therefore, in this case,
no relevant bound on $\Lambda_{\rm UV}$ can be obtained. Finally,
there are also scenarios for which the current experimental limits on
the generated NSI coefficients turn out to be too weak. An example of
this situation is shown in Fig.~\ref{fig:Plot5}, where we plot the NSI
coefficients obtained from the initial SMEFT WC
$C^{(1)}_{\protect\substack{\ell q \\ 1311}}$. The largest NSI
coefficients in this case are $\epsilon^{u L}_{e\tau}$ and
$\epsilon^{d L}_{e\tau}$, and these are only very weakly constrained.

In general, we have found that SMEFT scenarios with $\Lambda_{\rm UV}
> 3$ TeV lead to tiny NSI coefficients at low energies, always in
agreement with the current experimental bounds. For this reason, the
rest of the analysis will concentrate on NP scales between $0.5$ and
$3$ TeV. Moreover, the previous exploration allowed us to identify the
SMEFT scenarios capable to generate sizable NSI coefficients,
potentially resulting in relevant limits on the SMEFT WCs. Obtaining
these limits is our next goal.

\subsection{Limits on SMEFT Wilson coefficients from neutrino NSI}
\label{subsec:limits}

After our previous exploration, we have identified $18$ SMEFT scenarios
that give sizeable NSI coefficients for NP scales $\gtrsim
1\ \text{TeV}$ and performed again the procedure described in
Sec.~\ref{sec:NSIsEFTs} for each of them. After evaluating the
resulting NSI coefficients at low energies assuming $|C_i| = 1$ at $\mu
= \Lambda_{\rm UV}$ for different $\Lambda_{\rm UV}$ values, one can
easily interpolate to determine the value of the NP scale that
corresponds to the experimental bound of the NSI coefficient, thus
setting in this way a lower bound on the combination $\Lambda_{\rm
  UV}/\sqrt{|C_i|}$ for that particular SMEFT WC.

The bounds obtained with this method range between $\sim 700$ GeV and
$\sim 2.1$ TeV. We observe that the flavor violating operators
involving the 2nd and 3rd lepton generations get the strongest
bounds. This is because the most restrictive neutrino NSI experimental
bound, compiled in Ref.~\cite{Farzan:2017xzy}, is the NC NSI involving
quarks and the muon and tau neutrinos. This bound, derived from
IceCube DeepCore data~\cite{Aartsen:2017xtt}, sets the lower limit of
$\Lambda_{\rm UV}/\sqrt{|C_i|}$ for three different SMEFT WCs above
$\sim 2$ TeV. One can in principle find stronger bounds for the
associated NSI coefficients in \cite{Salvado:2016uqu}. However, the
range of neutrino energies used to derive these limits is mostly above
the electroweak scale and therefore cannot be used in our analysis.

We compare the constraining power of neutrino NSI with that of other
experimental signatures in the SMEFT. We consider two classes of WCs:

\begin{itemize}
\item {\bf Lepton Flavor Violating (LFV) coefficients:} The lack of
  signals of charged lepton flavor violating (CLFV) processes is known
  to strongly constrain the parameter space of many NP scenarios. This
  is expected to hold also for the SMEFT. Here we consider the
  radiative decays $\mu\to e \gamma$, $\tau\to e \gamma$ and $\tau\to
  \mu \gamma$, explored in detail in the context of the
  LEFT~\cite{Dekens:2018pbu}. For each of the LFV scenarios considered
  in our phenomenological analysis we derive a limit on the SMEFT
  combination $\Lambda_{\rm UV}/\sqrt{|C_i|}$. This is achieved with
  the same method as for neutrino NSI: for several values of
  $\Lambda_{\rm UV}$, the RGEs are evaluated down to the electroweak
  scale, where the SMEFT and LEFT are matched at tree-level, and then
  we further run down to $\Lambda_{\rm IR}$, where we impose the
  current 90\% C.L. bounds on the branching ratios of these
  processes~\cite{Baldini:2013ke,TheMEG:2016wtm,Aubert:2009ag} to
  determine bounds on the LEFT WCs. This indirectly translates into
  bounds on the SMEFT, in exactly the same way neutrino NSI
  experimental bounds do. This way, we obtain limits for two LFV
  SMEFT WCs. We also compute the $Z\to\tau\mu$ branching ratio  as 
  a function of $C_{\substack{H \ell \\23}}^{(1)}$ and $C_{\substack{H \ell \\23}}^{(3)}$ 
  using the effective coupling in Eq.~\eqref{eq:Zcoupling} and then compare it to the 
  experimental limit from~\cite{Aad:2016blu} to set a bound on the size of these WCs.
  In addition, we take the bounds compiled in Ref.~\cite{Carpentier:2010ue} for 
  $C_{\substack{\ell q \\1211}}^{(1)}$, $C_{\substack{\ell q \\2311}}^{(1)}$, 
  $C_{\substack{\ell q \\2311}}^{(3)}$,  $C_{\substack{\ell u \\2311}}$
  and $C_{\substack{\ell d \\2311}}$ .
  As expected, the limits on the SMEFT WCs derived from
  $\mu-e$ flavor violating processes are much stronger than those from
  $\tau-\mu$ or $\tau-e$ processes.
\item {\bf Lepton Flavor Conserving (LFC) coefficients:}
  Refs.~\cite{Falkowski:2015krw,Falkowski:2017pss} compile bounds on
  flavor conserving 4-fermion SMEFT operators derived from a plethora
  of low-energy experiments. The list includes lepton colliders,
  neutrino scattering on electron or nucleon targets, atomic parity
  violation, parity-violating electron scattering, as well as several
  precisely measured decays. For the coefficient $C_{\substack{H \ell \\11}}^{(3)}$ 
  we extract the bound from~\cite{Ellis:2018gqa} and for $C_{\substack{\ell q \\3311}}^{(3)}$ 
  we use LHC ditau measurements~\cite{Cerri:2018ypt}. The authors of these references present
  their results in the form of 68\% C.L. ranges for the coefficients
  the LFC operators. Assuming a Gaussian distribution, and taking into
  account some minor differences in notation and conventions, we
  \emph{translate} these ranges into 90\% C.L. bounds in order to have
  a fair comparison to the bounds derived from neutrino NSI. We get
  limits on $\Lambda_{\rm UV}/\sqrt{|C_i|}$ for seven LFC WCs.
\end{itemize} 	
 
{
\begin{table}[t!]
  \centering
  \begin{tabularx}{\textwidth}{ >{\setlength\hsize{0.6\hsize}\centering}X>{\setlength\hsize{0.7\hsize}\centering}X>{\setlength\hsize{1.2\hsize}\centering}X >{\setlength\hsize{1.3\hsize}\centering}X >{\setlength\hsize{1.2\hsize}}X } 
    
    \hline\hline
    SMEFT & NSI  &&Other& \tabularnewline	\hline	 &Coefficient  & $\Lambda_{\rm UV}/\sqrt{|C_i|}$ (TeV)& Process & $\Lambda_{\rm UV}/\sqrt{|C_i|}$ (TeV) \tabularnewline\hline\hline

    \rule{0pt}{15pt}	$C_{\substack{H \ell \\12}}^{(1)}$ & $\epsilon^{u L}_{e\mu}$ & $>0.91$  & $\mu\to e\gamma$  & $>6.10$   \tabularnewline
    \rule{0pt}{15pt}	$C_{\substack{H \ell \\23}}^{(1)}$ & $\epsilon^{d V}_{\mu\tau}$ & $>1.53$ &  $Z\to\mu \tau$  & $>1.72$   \tabularnewline
    \rule{0pt}{15pt}	$C_{\substack{H \ell \\11}}^{(3)}$ & $\epsilon^{e L}_{ee}$ & $>1.13$  & LHC \cite{Ellis:2018gqa} & $>4.89$  \tabularnewline  
    \rule{0pt}{15pt}	$C_{\substack{H \ell \\23}}^{(3)}$ & $\epsilon^{dV}_{\mu\tau}$ & $>1.72$   &  $Z\to\mu \tau$  & $>1.72$  \tabularnewline
    \rule{0pt}{15pt}	$C_{\substack{\ell \ell \\2211}}$ & $\epsilon^{e L}_{\mu\mu}$ & $>1.43$ & \cite{Falkowski:2017pss,Falkowski:2015krw} &$>3.29$   \tabularnewline
    \rule{0pt}{15pt}	$C_{\substack{\ell q \\1111}}^{(1)}$ & $\epsilon^{uV}_{ee}$ & $>0.99$  & \cite{Falkowski:2017pss,Falkowski:2015krw} & $>4.58$  \tabularnewline
    \rule{0pt}{15pt}	$C_{\substack{\ell q \\1211}}^{(1)}$ & $\epsilon^{uL}_{e\mu}$ & $>1.11$  & $\mu \to e$ in Ti \cite{Carpentier:2010ue}  &$>267.06$  \tabularnewline
    \rule{0pt}{15pt}	$C_{\substack{\ell q \\2311}}^{(1)}$ & $\epsilon^{dV}_{\mu\tau}$ & $>2.10$ &  $\tau$ decays \cite{Carpentier:2010ue}  &$>7.87$  \tabularnewline
    \rule{0pt}{15pt}	$C_{\substack{\ell q \\1111}}^{(3)}$ & $\epsilon^{uV}_{ee}$ & $>1.00$ &\cite{Falkowski:2017pss,Falkowski:2015krw}& $>8.65$\tabularnewline
    \rule{0pt}{15pt}	$C_{\substack{\ell q \\1211}}^{(3)}$ & $\epsilon^{uL}_{e\mu}$ & $>1.10$  & $\mu\to e\gamma$  & $>14.97$  \tabularnewline
    \rule{0pt}{15pt}	$C_{\substack{\ell q \\2311}}^{(3)}$ & $\epsilon^{dV}_{\mu\tau}$ & $>1.99$&   $\tau$ decays \cite{Carpentier:2010ue}  &$>7.87$  \tabularnewline
    \rule{0pt}{15pt}	$C_{\substack{\ell q \\3311}}^{(3)}$ & $\epsilon^{dV}_{\tau\tau}$ & $>0.93$ & LHC \cite{Cerri:2018ypt} &$>4.67$\tabularnewline
    \rule{0pt}{15pt}	$C_{\substack{\ell e \\2211}}$ & $\epsilon^{eR}_{\mu\mu}$ & $>1.01$ & \cite{Falkowski:2017pss,Falkowski:2015krw}& $>2.90$ \tabularnewline
    \rule{0pt}{15pt}	$C_{\substack{\ell u \\1111}}$ & $\epsilon^{uV}_{ee}$ & $>1.03$ &\cite{Falkowski:2017pss,Falkowski:2015krw}&$>3.32$ \tabularnewline
    \rule{0pt}{15pt}	$C_{\substack{\ell u \\2311}}$ & $\epsilon^{uV}_{\mu\tau}$ & $>2.10$ &   $\tau$ decays \cite{Carpentier:2010ue}  &$>7.87$  \tabularnewline
    \rule{0pt}{15pt}	$C_{\substack{\ell d \\1111}}$ & $\epsilon^{dV}_{ee}$ & $>1.01$ &\cite{Falkowski:2017pss,Falkowski:2015krw}& $>3.12$\tabularnewline
    \rule{0pt}{15pt}	$C_{\substack{\ell d \\2211}}$ & $\epsilon^{dV}_{\mu\mu}$ & $>0.69$ &\cite{Falkowski:2017pss,Falkowski:2015krw}&$>0.94$ \tabularnewline
    \rule{0pt}{15pt}	$C_{\substack{\ell d \\2311}}$ & $\epsilon^{dV}_{\mu\tau}$ & $>2.13$ &   $\tau$ decays \cite{Carpentier:2010ue}  &$>7.87$ \rule{0pt}{15pt}  \tabularnewline \tabularnewline 
    \hline						
  \end{tabularx}
  \caption{Lower limits on $\Lambda_{\rm UV}/\sqrt{|C_i|}$, with
    $\Lambda_{\rm UV}$ the NP scale and $C_i$ the SMEFT WC. This
    tables compares the limits derived from neutrino NSI with the
    limits obtained from other experimental signatures: LFV processes
    as well as collider experiments and low-energy
    LFC measurements. 
\label{ta:Lambda}
}
\end{table}
}

\begin{figure}[t!]
  \centering
  \includegraphics[width=0.7\linewidth]{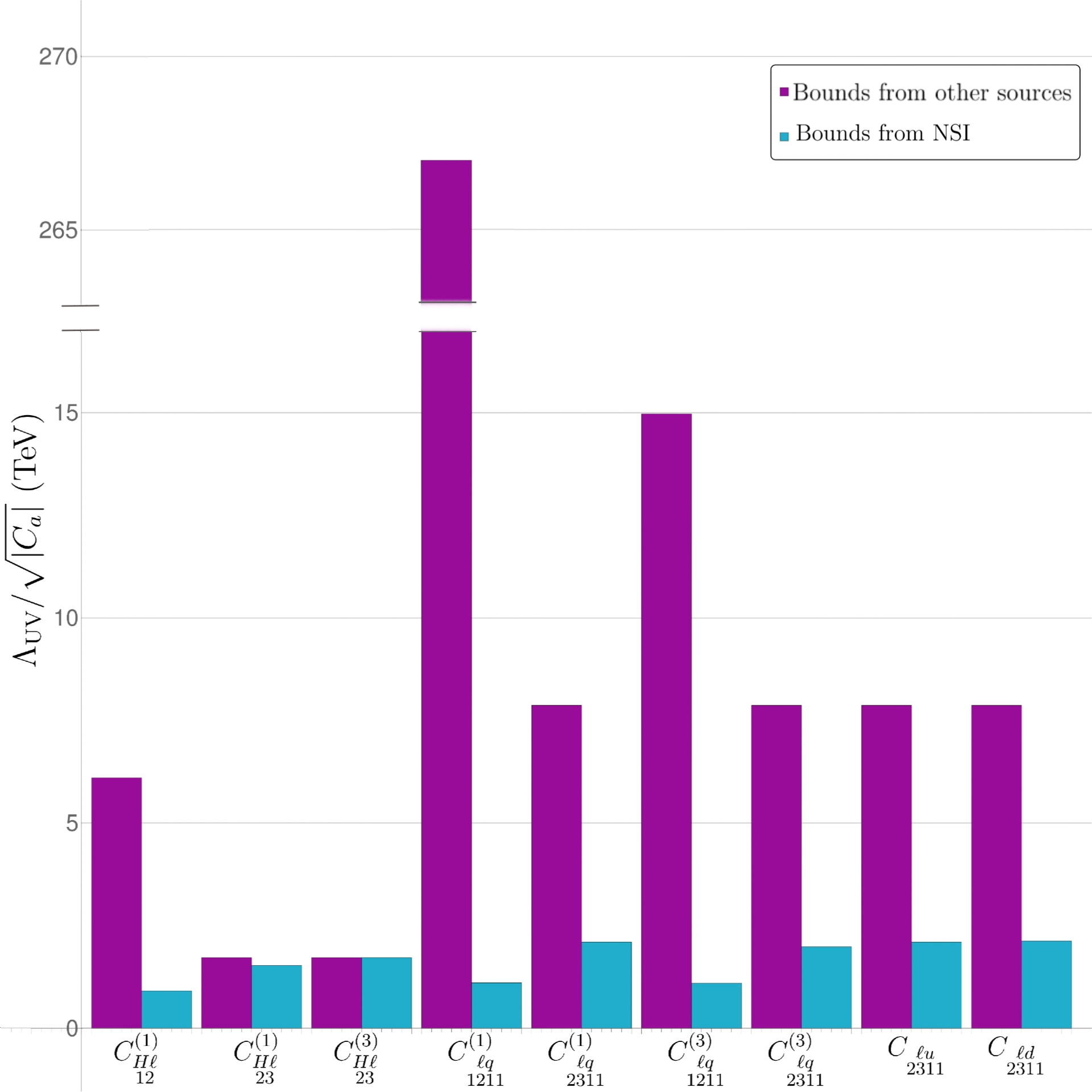}
  \caption{Lower limits on $\Lambda_{\rm UV}/\sqrt{|C_i|}$, with
    $\Lambda_{\rm UV}$ the NP scale and $C_i$ the SMEFT WC, for
    several LFV SMEFT WCs, derived from neutrino NSI (blue bars) and
    from LFV processes (purple bars). See text for details.
    \label{fig:limitsLFV}
  }
\end{figure}

\begin{figure}[t!]
  \centering
  \includegraphics[width=0.7\linewidth]{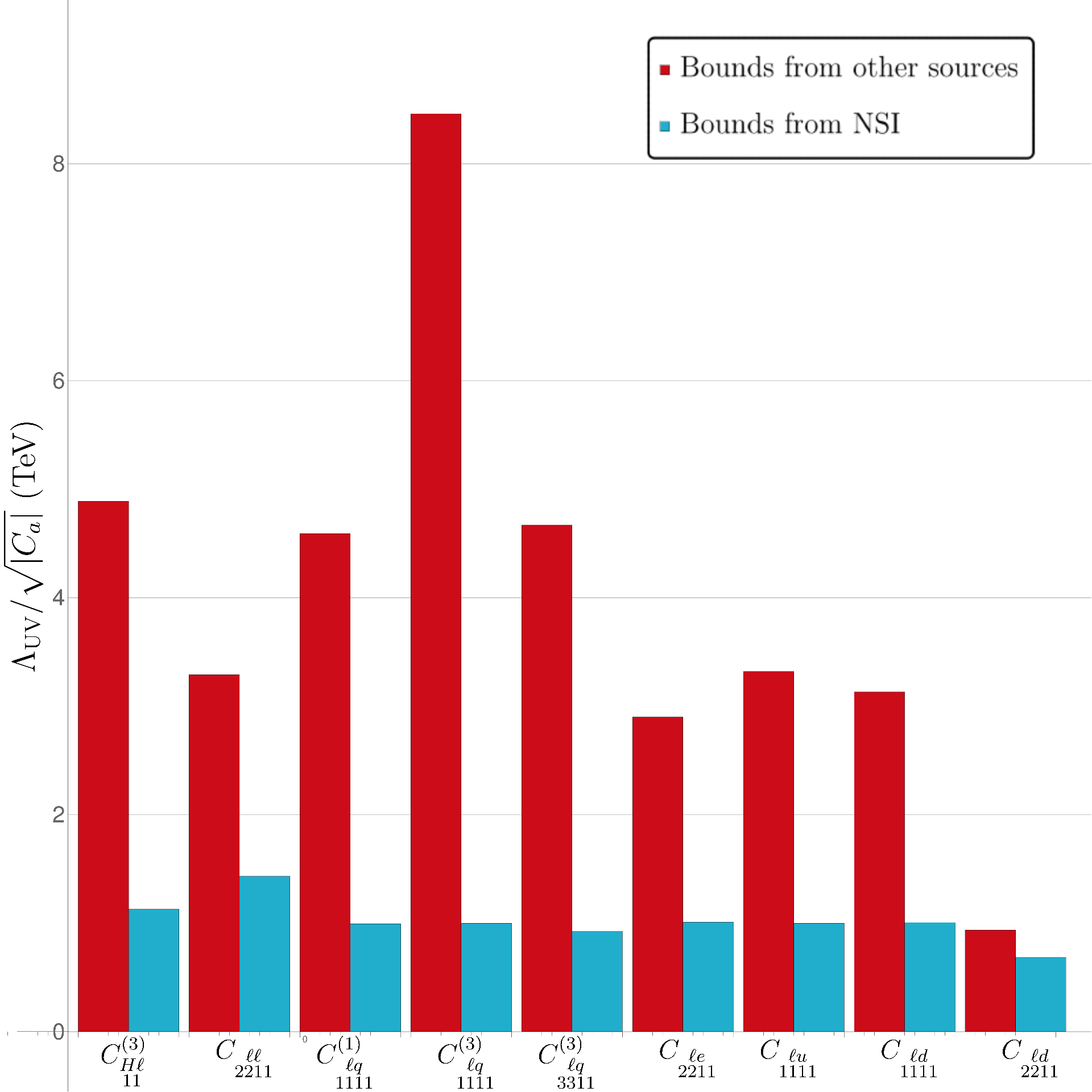}
  \caption{Lower limits on $\Lambda_{\rm UV}/\sqrt{|C_i|}$, with
    $\Lambda_{\rm UV}$ the NP scale and $C_i$ the SMEFT WC, for
    several LFC SMEFT WCs, derived from neutrino NSI (blue bars) and
    from  several LFC processes (red bars).
    \label{fig:limitsLFC}
  }
\end{figure}

Our results are presented in Figs.~\ref{fig:limitsLFV} and
\ref{fig:limitsLFC} and compiled in Tab.~\ref{ta:Lambda}. As
anticipated, the limits from CLFV decays are more stringent than those
from neutrino NSI. In fact, in some cases, the NP scale is constrained
to be above $\sim 15$ TeV for $\mathcal{O}(1)$ WCs. Also, the bounds
obtained from low-energy precision measurements (and extracted from
\cite{Falkowski:2015krw,Falkowski:2017pss}) are typically more
constraining than those derived from neutrino NSI.

We point out that the weakest limit obtained with neutrino NSI
experiments is for the $C_{\substack{\ell d \\2211}}$ coefficient,
restricted to $\Lambda_{\rm UV}/\sqrt{|C_{\substack{\ell d \\2211}}|}
> 690$ GeV. Since $690$ GeV is well above the electroweak scale, we
consider the SMEFT approach justified. In any case, the bound from
low-energy precision
measurements~\cite{Falkowski:2015krw,Falkowski:2017pss} is stronger,
pushing $\Lambda_{\rm UV}$ to almost the TeV scale and, therefore, the
potential NP degrees of freedom contributing to $C_{\substack{\ell d
    \\2211}}$ would in principle be even heavier.

Finally, a comment on the relevance of RGE running is in order. Our
previous analysis includes the full one-loop RGE running in the SMEFT
and the LEFT implemented in \dsix~\cite{Celis:2017hod,dsixv}. It is
therefore essential to assess the importance of the RGE running in our
procedure. To do so, we have repeated the process described in
Sec.~\ref{sec:NSIsEFTs} and computed the NSI coefficients resulting
from the different SMEFT scenarios considered in our analysis, this
time without RGE running. Comparing to our previous results
one can evaluate the relevance of running effects. For $\Lambda_{\rm
  UV} = 1$ TeV, the relative difference in the resulting NSI
coefficients lies between 5\% and 10\%. This difference grows, as
expected, for higher NP scales. One can also determine the impact on
the bounds on $\Lambda_{\rm UV}/\sqrt{|C_i|}$. For example, in the
SMEFT scenario with a non-vanishing $C_{\substack{H \ell
    \\23}}^{(1)}$, the derived bound changes from $1.53$ TeV to $1.61$
TeV. A similar change takes place in case of $C_{\substack{\ell q
    \\2311}}^{(3)}$, which goes from $1.99$ TeV when RGE running is
included to $1.89$ TeV when it is absent. Therefore, although a
numerical change can be noticed in some cases, the global picture
would not be affected if running effects are neglected. Nevertheless,
we emphasize that this conclusion holds for the scenarios considered
in our exploratory analysis. One cannot discard more relevant running
effects in other regions of the vast SMEFT parameter space.

\section{Summary and discussion}
\label{sec:conclusions}

Neutrino NSI constitute a powerful method to constrain NP at low
energies. However, due to the absence of direct experimental evidence
of their existence, the NP degrees of freedom might actually lie at
very high energies, clearly above the electroweak scale. In this paper,
we bridge the energy gap between the experiments setting limits on the
neutrino NSI coefficients and the parameters of the SMEFT, an EFT
valid at high energies. This connection allows for an easy application
of our results to a very general class of NP models.

Our main results are shown in Figures~\ref{fig:limitsLFV} and
\ref{fig:limitsLFC} and compiled in Table~\ref{ta:Lambda}. We conclude
that current NSI limits already push the NP scale above the TeV in
most cases. We also find that limits from other experimental probes,
in particular from low-energy measurements or lepton flavor violating
searches, are stronger and require higher values for $\Lambda_{\rm
  UV}$. While the results obtained in our analysis lead to the same
qualitative conclusions reached by previous
works~\cite{Altmannshofer:2018xyo,Falkowski:2019xoe,Davidson:2019iqh},
we emphasize the inclusion of full one-loop RGE running effects at low
and high energies. This has allowed us to derive robust bounds on the
SMEFT WCs and assess the numerical relevance of the running effects in
the scenarios we have considered.

There are several ways in which our analysis can be extended. It is
well-known that dimension-8 operators may play a relevant role, see
for instance \cite{Altmannshofer:2018xyo}. One should also bear in
mind that our analysis assumes one SMEFT WC at a time. In more general
scenarios cancellations are in principle expected, potentially
weakening the bounds. Similarly, we note that many of the SMEFT
operators considered in our analysis generate both NC-NSI and CC-NSI,
and therefore these should be taken into account simultaneously in
order to derive consistent constraints. Finally, one can extend the
SMEFT with additional fields. For instance, operators involving light
sterile neutrinos, singlet under the SM gauge group, have been
considered in several
works~\cite{delAguila:2008ir,Aparici:2009fh,Alonso:2014zka,Bhattacharya:2015vja,Liao:2016qyd,Alcaide:2019pnf}. They
allow for new scalar and tensorial neutrino four-fermion interactions
at low energies~\cite{Lindner:2016wff,AristizabalSierra:2018eqm},
recently shown to offer new phenomenological possibilities of
interest~\cite{Bischer:2018zcz,Bischer:2019ttk}.

\section*{Acknowledgements}

The authors are grateful to Valentina De Romeri, Martin Jung and
Christoph A. Ternes for many enlightening discussions and useful
comments. AV is grateful to Mart\'in Gonz\'alez-Alonso for fruitful
discussions and clarifications about
Refs.~\cite{Falkowski:2015krw,Falkowski:2017pss}. JTC is thankful to MT and AV for their great support in the beginning of his academic career.
Work supported by the Spanish grants FPA2017-85216-P
(MINECO/AEI/FEDER, UE), SEJI/2018/033 and PROMETEO/2018/165 grants
(Generalitat Valenciana), and FPA2017-90566-REDC (Red Consolider
MultiDark).
MT acknowledges financial support from MINECO through the Ram\'{o}n y Cajal contract RYC-2013-12438.

\appendix

\section{SMEFT in the broken phase}
\label{sec:app1}

Several SM parameters get modified after eletroweak symmetry breaking
due contributions from dimension-6 SMEFT operators. We compile in this
Appendix their explicit analytical expressions. These definitions have
been extracted from \cite{Jenkins:2017jig}.

The Higgs vacuum expectation value is modified by the operator
$\Q_H=C_H\left(H^\dagger H\right)^3$, which describes a
6-Higgs interaction, as
\begin{equation}\label{eq:DefI} 
    v_T\equiv \left(1+\dfrac{3 \, C_H \, v^2}{8 \, \lambda \, \Lambda_{\rm UV}^2}\right) v \, .
\end{equation}
We define the expansion parameter
\begin{equation}
\delta_T\equiv \dfrac{v_T}{\Lambda_{\rm UV}} \, (\ll 1) \, .
\end{equation}
The gauge couplings, the weak mixing angle and the effective photon
and $Z$-boson couplings get also modified by dimension-6 SMEFT operators
involving the SM gauge fields and the Higgs doublet. They are given by
\begin{equation}
\Bar{g}_1=g_1\left(1+C_{HB} \, \delta_T^2\right),\qquad\Bar{g}_2=g_2\left(1+C_{HW} \, \delta_T^2\right) \, .
\end{equation}
\begin{align}
\cos \Bar{\theta}\equiv\Bar{c}=\dfrac{\Bar{g}_2}{\sqrt{\Bar{g}_1^2+\Bar{g}_2^2}}\left[1-\delta_T^2 \, \dfrac{C_{HWB}}{2}\dfrac{\Bar{g}_1}{\Bar{g}_2}\left(\dfrac{\Bar{g}_2^2-\Bar{g}_1^2}{\Bar{g}_2^2+\Bar{g}_1^2}\right)\right] \, ,\nonumber\\
\sin \Bar{\theta}\equiv\Bar{s}=\dfrac{\Bar{g}_1}{\sqrt{\Bar{g}_1^2+\Bar{g}_2^2}} \left[1+\delta_T^2 \, \dfrac{C_{HWB}}{2}\dfrac{\Bar{g}_2}{\Bar{g}_1}\left(\dfrac{\Bar{g}_2^2-\Bar{g}_1^2}{\Bar{g}_2^2+\Bar{g}_1^2}\right)\right] \, ,
\end{align}
and
\begin{equation}
\Bar{e}=\Bar{g}_2\Bar{s}-\dfrac{1}{2}\Bar{c} \, \Bar{g}_2\, \delta_T^2 \, C_{HWB},\qquad \Bar{g}_Z^2=\dfrac{\Bar{e}}{\Bar{s}\Bar{c}}\left[1+\dfrac{\Bar{g}_1^2+\Bar{g}_2^2}{2\Bar{g}_1 \, \Bar{g}_2} \, \delta_T^2 \, C_{HWB} \right] \, .
\end{equation}
Finally, the fermion couplings to the $W$ and $Z$ bosons are also
modified by dimension-6 SMEFT operators. The $W$-boson couplings are
given by
\begin{equation}
  \left[W_\ell\right]_{pr}=\left[\delta_{pr}+\, \delta_T^2 \, C_{\substack{H\ell\\pr}}^{(3)}\right] \, , \quad \left[W_q\right]_{pr}=\left[\delta_{pr}+\, \delta_T^2 \, C_{\substack{H q\\pr}}^{(3)}\right] \, , \quad \left[W_R\right]_{pr}=\left[\dfrac{1}{2}\, \delta_T^2 \, C_{\substack{Hud\\pr}}\right] \, ,
\end{equation}
while the $Z$-boson couplings can be expressed as
\begin{align}
  \left[Z_\nu\right]_{pr}=&\left[\dfrac{1}{2}\delta_{pr}-\dfrac{1}{2}\, \delta_T^2\, C_{\substack{H\ell\\pr}}^{(1)}+\dfrac{1}{2}\, \delta_T^2 \, C_{\substack{H\ell\\pr}}^{(3)}\right]\, , \\
  \left[Z_{e_L}\right]_{pr}=&\left[\delta_{pr}\left(-\dfrac{1}{2}+\Bar{s}^2\right)-\dfrac{1}{2}\, \delta_T^2 \, C_{\substack{H\ell\\pr}}^{(1)}-\dfrac{1}{2}\, \delta_T^2 \, C_{\substack{H\ell\\pr}}^{(3)}\right] \, ,\label{eq:Zcoupling} \\
  \left[Z_{e_R}\right]_{pr}=&\left[\delta_{pr}\Bar{s}^2-\dfrac{1}{2}\, \delta_T^2 \, C_{\substack{He\\pr}}\right] \, , \\
  \left[Z_{u_L}\right]_{pr}=&\left[\delta_{pr}\left(\dfrac{1}{2}-\dfrac{2}{3}\Bar{s}^2\right)-\dfrac{1}{2}\, \delta_T^2 \, C_{\substack{Hq\\pr}}^{(1)}+\dfrac{1}{2}\, \delta_T^2 \, C_{\substack{Hq\\pr}}^{(3)}\right] \, , \\
  \left[Z_{u_R}\right]_{pr}=&\left[\delta_{pr}\left(-\dfrac{2}{3}\Bar{s}^2\right)-\dfrac{1}{2}\, \delta_T^2 \, C_{\substack{Hu\\pr}}\right] \, , \\
  \left[Z_{d_L}\right]_{pr}=&\left[\delta_{pr}\left(-\dfrac{1}{2}+\dfrac{1}{3}\Bar{s}^2\right)-\dfrac{1}{2}\, \delta_T^2 \, C_{\substack{Hq\\pr}}^{(1)}-\dfrac{1}{2}\, \delta_T^2 \, C_{\substack{Hq\\pr}}^{(3)}\right] \, , \\
  \left[Z_{d_R}\right]_{pr}=&\left[\delta_{pr}\dfrac{1}{3}\Bar{s}^2-\dfrac{1}{2}\, \delta_T^2 \, C_{\substack{Hd\\pr}}\right] \, .
\end{align}

\providecommand{\href}[2]{#2}\begingroup\raggedright\endgroup

\end{document}